\documentclass[a4paper,preprint]{article}

\usepackage{18lomcon}        
\usepackage{cite}             
\usepackage{epsfig}           
\usepackage{float}
\usepackage{amsmath}
\usepackage{url}
\usepackage{subcaption}
\newcommand{\bra}[1]{\langle #1|} 
\newcommand{\ket}[1]{|#1\rangle}

\begin{document}
\null\hfill\begin{tabular}[t]{l@{}}
  HIP-2017-24/TH \\
  \\
\end{tabular}
\title{PURSUIT FOR OPTIMAL BASELINE FOR MATTER NONSTANDARD INTERACTIONS IN LONG BASELINE NEUTRINO OSCILLATION EXPERIMENTS}

\author{Timo K{\"a}rkk{\"a}inen \email{timo.j.karkkainen@helsinki.fi}
}

\affiliation{Department of Physics, University of Helsinki, FI-00014 Helsinki, Finland}

\date{}
\maketitle
\begin{abstract}
We investigate the prospects for probing the strength of the possible matter nonstandard neutrino interactions (mNSI) in long baseline neutrino oscillation experiments and the interference of the leptonic CP angle $\delta_{CP}$ with the constraining of the mNSI couplings. The interference is found to be strong in the case of the $\nu_e \leftrightarrow \nu_\mu$ and $\nu_e \leftrightarrow \nu_\tau$ transitions but not significant in the other cases.
\end{abstract}

\section{Introduction}

After the confirmation of atmospheric and solar oscillations, the neutrino physicist community has fitted the standard three flavor neutrino oscillations (NO) framework to the continuously accumulating data. As the mixing parameters are determined more and more precisely, it is clear that neutrino flavor transition can be very well interpreted as a purely oscillatory phenomenon. However, there may be subleading contributions to flavor transition. In that scenario, the future datasets will turn out to be unfittable to the NO framework, and additional degrees of freedom must be introduced.

\section{Formalism}

The possible deviations of standard NO scheme may be parametrized by nonstandard interactions (NSI). Consider the following Lagrangians, which are nonrenormalizable and not gauge invariant:
\begin{align}\label{lnsi}
\mathcal{L}^\text{CC}_\text{NSI} &= -2\sqrt{2}G_F\varepsilon_{\alpha\beta}^{ff',C}(\overline{\nu}_{\alpha}\gamma^{\mu}P_L\ell_{\beta})(\overline{f}\gamma^{\mu}P_Cf'),\: f \neq f', \notag \\
\mathcal{L}^\text{NC}_\text{NSI} &= -2\sqrt{2}G_F\varepsilon_{\alpha\beta}^{f,C}(\overline{\nu}_{\alpha}\gamma^{\mu}P_L\nu_{\beta})(\overline{f}\gamma^{\mu}P_Cf).
\end{align}
In the most general case, $f$ and $f'$ are charged leptons or quarks, $G_F$ is the Fermi coupling constant, $\alpha$ and $\beta$ are flavor labels, and $C$ is the chiral label, $P_L$ and $P_R$ being the chiral projection operators. The NSI parameters $\varepsilon_{\alpha\beta}^{ff',C}$ and $\varepsilon_{\alpha\beta}^{f,C}$ are dimensionless complex numbers. The absolute value of the number corresponds to relative strength of the new interaction with respect to Fermi interaction. This may be used to estimate the mass scale $M$ of the new interaction, with $\varepsilon \sim m_W^2/M$. The charged current Lagrangian $\mathcal{L}^\text{CC}_\text{NSI}$ is relevant for the NSI effects in the neutrino creation and detection processes and the neutral current Lagrangian $\mathcal{L}^\text{NC}_\text{NSI}$ is relevant for the NSI matter (mNSI) effects.  The effective low-energy NSI Lagrangians (\ref{lnsi}) are assumed to follow from some unspecified beyond-the-standard-model (BSM) theory after integrating out heavy degrees of freedom.

mNSI matrix elements are gained by summing over chirality and fermion states ($N_f$ and $N_e$ are the fermion $f$ and electron number densities, respectively),
\begin{equation}
\varepsilon^m_{\alpha\beta} =  \sum \limits_{f,C}\varepsilon_{\alpha\beta}^{fC}\frac{N_f}{N_e},
\end{equation}
Description of NO in matter is given by standard interaction (SI) Hamiltonian 
\begin{equation}
H_{\text{SI}} =\frac{1}{2E_{\nu}}\left[ U\text{diag}(m_1^2,m_2^2,m_3^2)U^{\dagger} + \left( \begin{array}{ccc}
V_{\text{CC}}+V_{\text{NC}} & 0 & 0\\
0 & V_{\text{NC}} & 0 \\
0 & 0 & V_{\text{NC}}
\end{array}\right) \right] ,
\end{equation}
where $m_{1,2,3}$ are neutrino masses, $U$ is neutrino mixing matrix, $E_\nu$ is neutrino energy, $V_{\text{CC}} = \sqrt{2}G_FE_{\nu}N_e$ and $V_{\text{NC}} = -\frac{\sqrt{2}}{2}G_FE_{\nu}N_n$ are the matter potentials. In the case of mNSI, NO probability is calculated with NSI Hamiltonian, which is the SI case with an extra term:
\begin{equation}
P(\nu_\alpha \rightarrow \nu_\beta) = \left|\bra{\nu_{\beta}}e^{-i(H_{\text{SI}}+ V_{\text{CC}}\varepsilon^m/2E_{\nu})L}\ket{\nu_{\alpha}}\right|^2.
\end{equation}
Note that we get the SI case at the $\varepsilon^m \rightarrow 0$ limit or when the density of matter is negligible, as expected.

\section{Numerical analysis and discussion}
We study how the future NO experiments would constrain various mNSI parameters, setting the baseline free and letting the NO parameters vary within their $3\sigma$ limits \cite{nufit}. We consider both mass hierarchies but only higher $\theta_{23}$ octant.

We use two benchmark setups, both designed for a future long baseline neutrino experiment with double-phase liquid argon detector. The first benchmark (\textbf{SPS}) utilized the LBNO setup with 20 kt detector and beam optimization at 2288 km\cite{art0}. Our second benchmark (\textbf{DUNE}) utilized the DUNE setup \cite{dune} with 40 kt detector and beam optimization at 1300 km. The analysis is done by using the GLoBES simulation software \cite{art}.

\begin{figure}[ht]
\centering
\includegraphics[width=12cm]{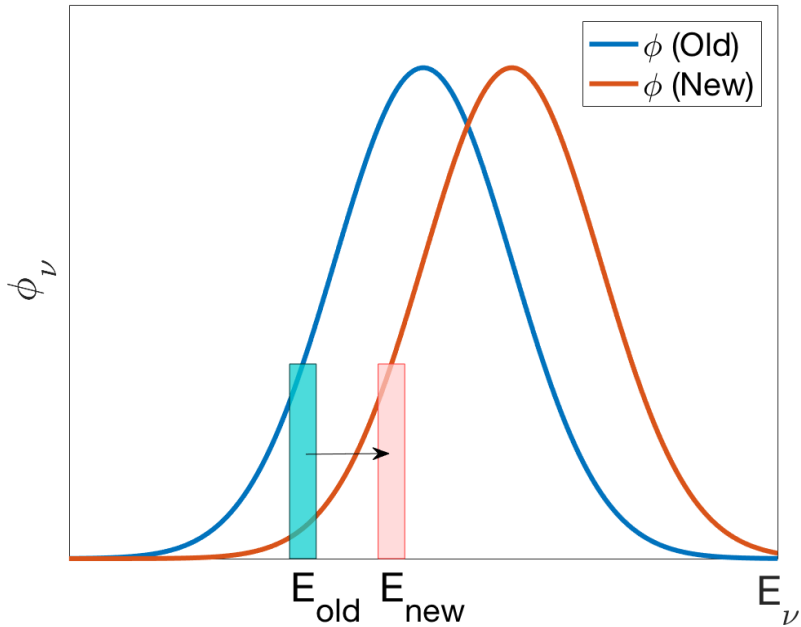}
\caption{\label{fig1} Shifting neutrino flux to preserve the baseline optimization of original. $E_\text{new}$ is calculated from Eq. (\ref{e}).}
\end{figure}

Because the neutrino fluxes are optimized at a certain baseline, once we perform the simulation with a different baseline, we reoptimize the flux assuming $L/E = $ constant, by
\begin{equation}\label{e}
E_\text{new} = \frac{L_\text{new}}{L_\text{old}}E_\text{old},\:\:\: \phi_\text{new}(E_\text{new}) = \phi_\text{old}(E_\text{old})
\end{equation}
See Figure~\ref{fig1} for a pictorial representation.

\begin{figure}[ht]
\centering
\begin{subfigure}{0.49\textwidth}
\includegraphics[width=6.0cm]{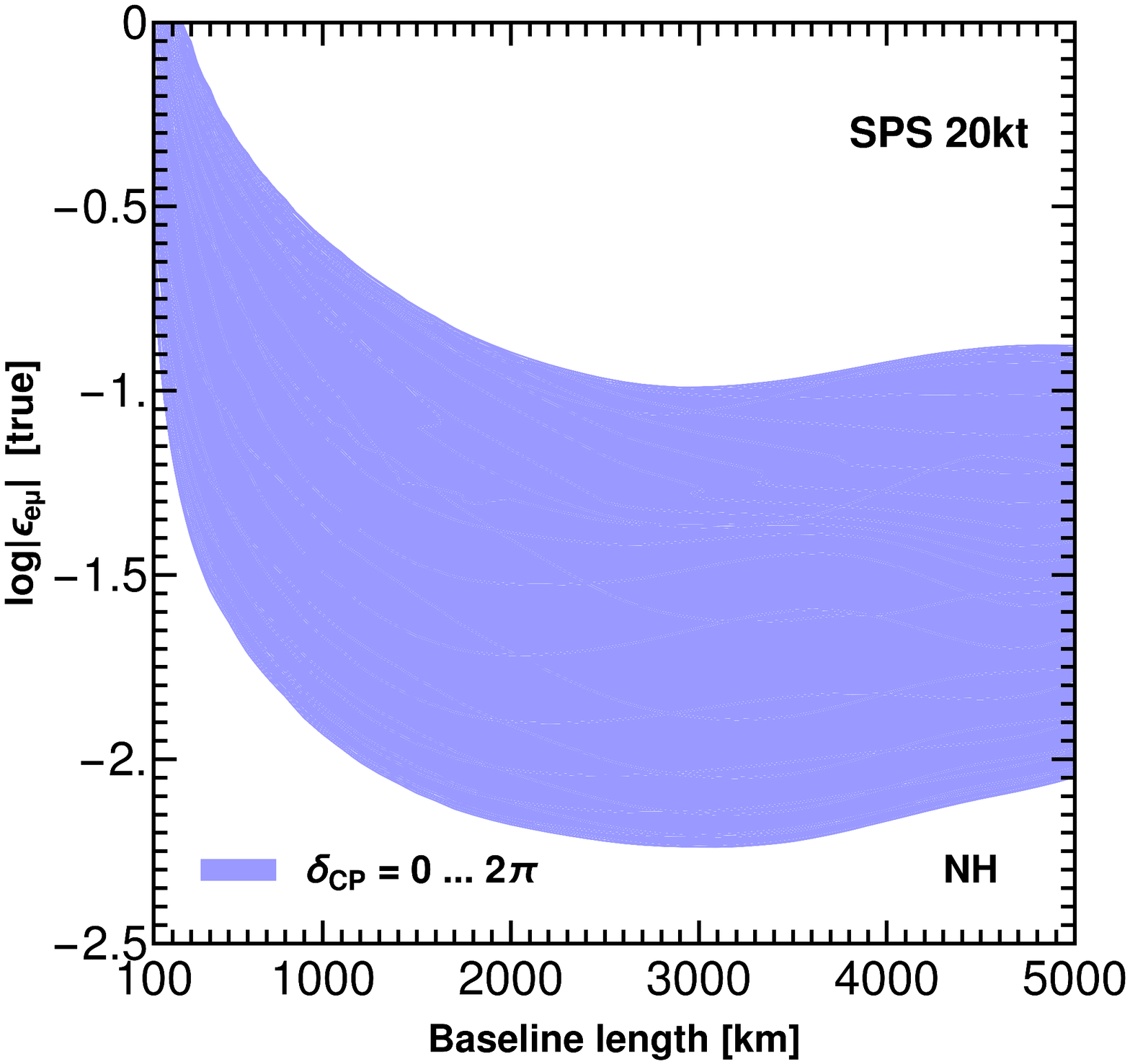}
\end{subfigure}
\begin{subfigure}{0.49\textwidth}
\includegraphics[width=6.0cm]{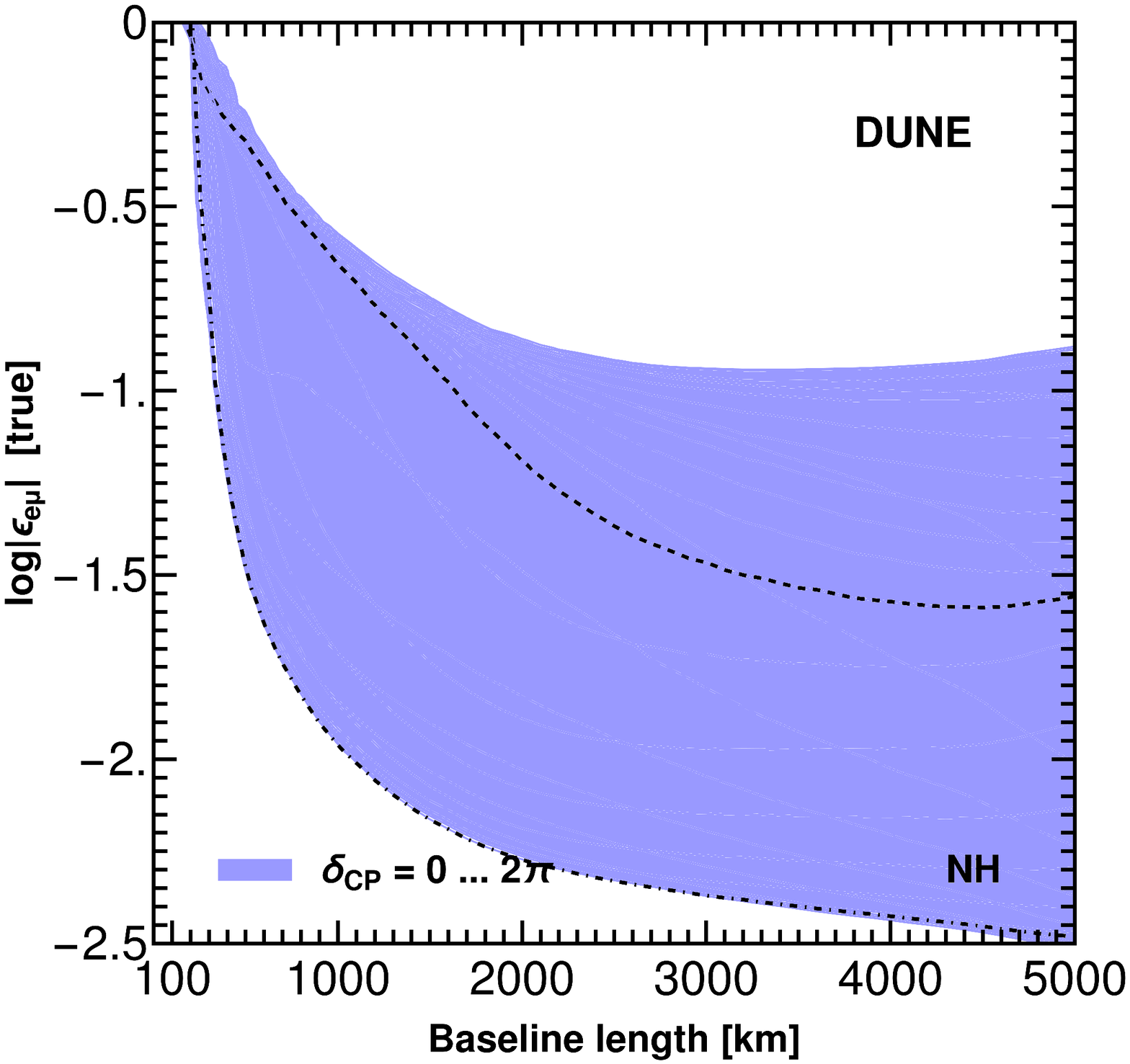}
\end{subfigure}
\begin{subfigure}{0.49\textwidth}
\includegraphics[width=6.0cm]{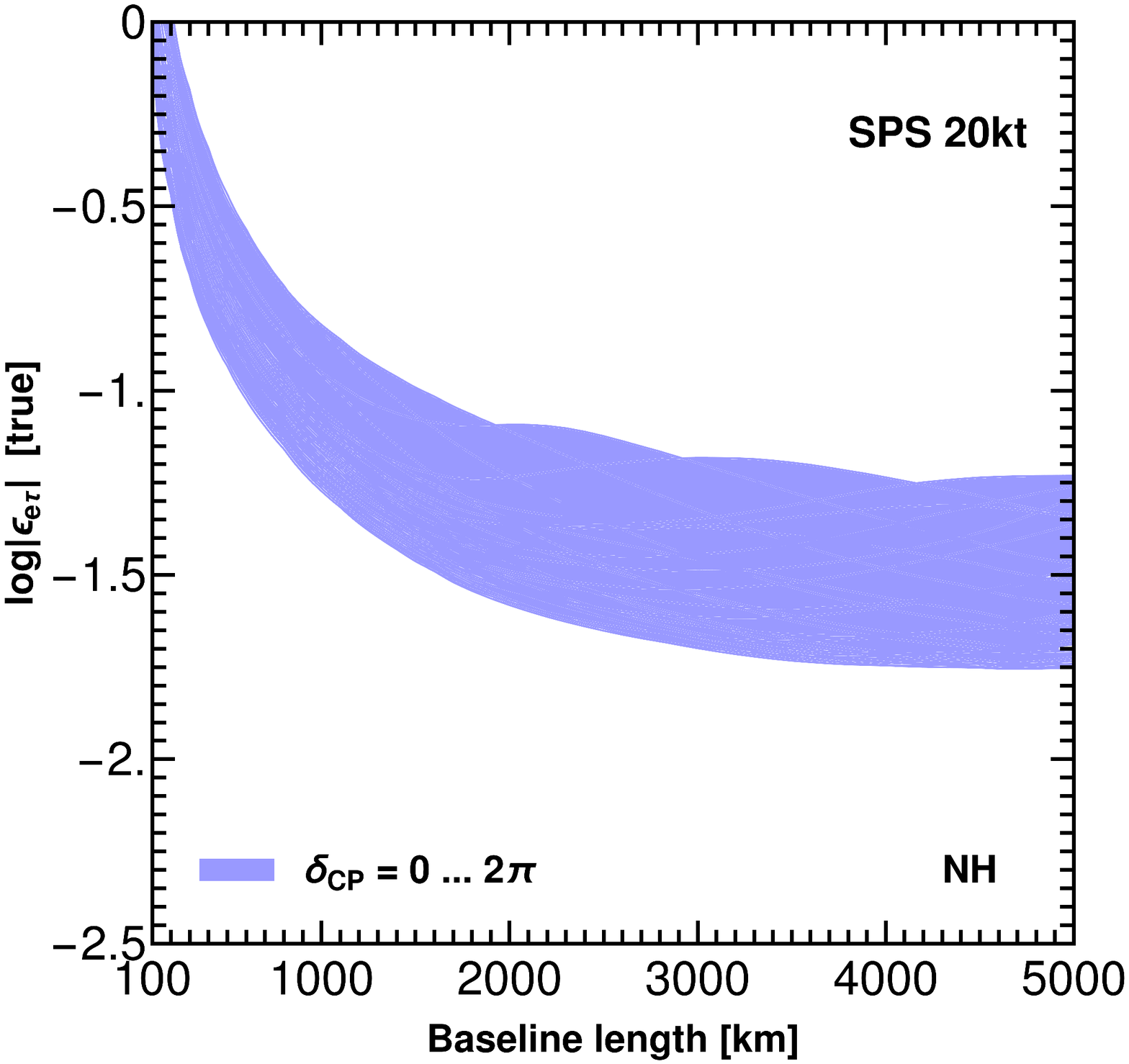}
\end{subfigure}
\begin{subfigure}{0.49\textwidth}
\includegraphics[width=6.0cm]{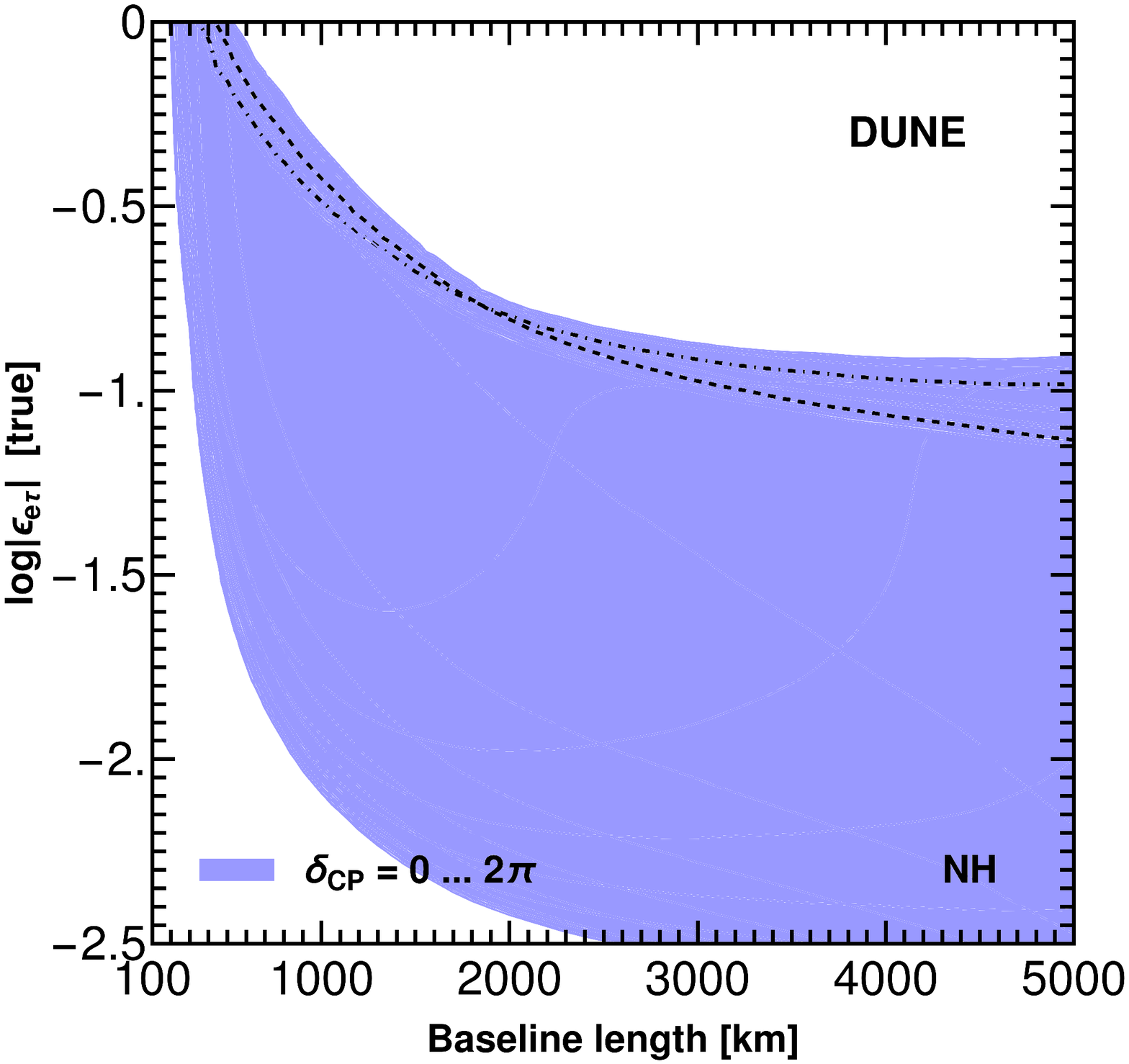}
\end{subfigure}
\caption{\label{fig2}90 $\%$ CL discovery reach of $|\varepsilon^m_{e\mu}|$ and $|\varepsilon^m_{e\tau}|$ as a function of baseline length for both benchmark setups. Vertical axes are logarithmic. Band thickness results from the ambiguity of $\delta_{CP}$, which visibly interferes with determining the upper bound for mNSI. Values above the band can be excluded with the given benchmark. Dashed line in DUNE plots represents the case $\delta_{CP} = 0$ and dot-dashed line the case $\delta_{CP} = \pi/2$.}
\end{figure}

\begin{figure}[ht]
\centering
\begin{subfigure}{0.49\textwidth}
\includegraphics[width=6.0cm]{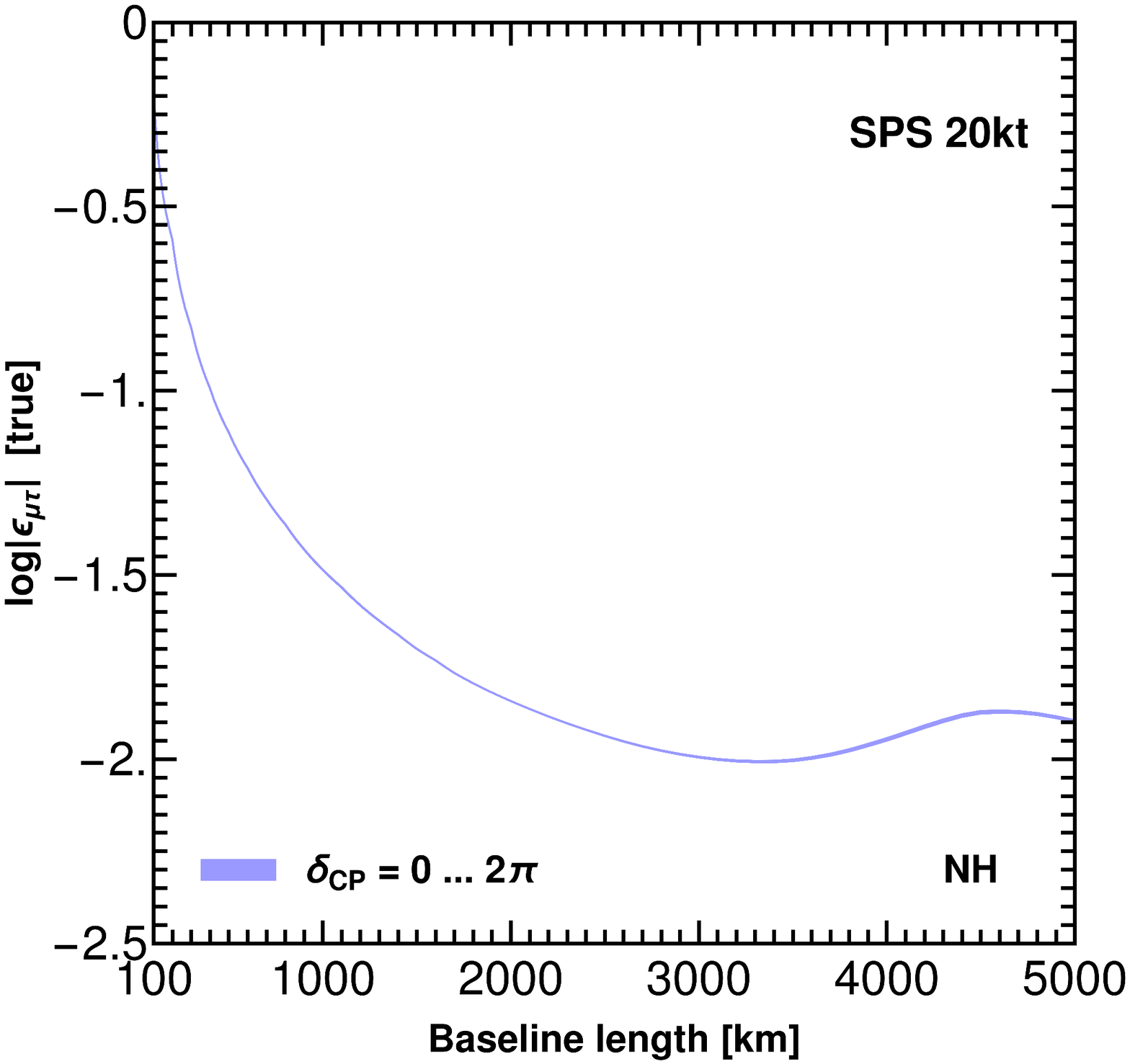}
\end{subfigure}
\begin{subfigure}{0.49\textwidth}
\includegraphics[width=6.0cm]{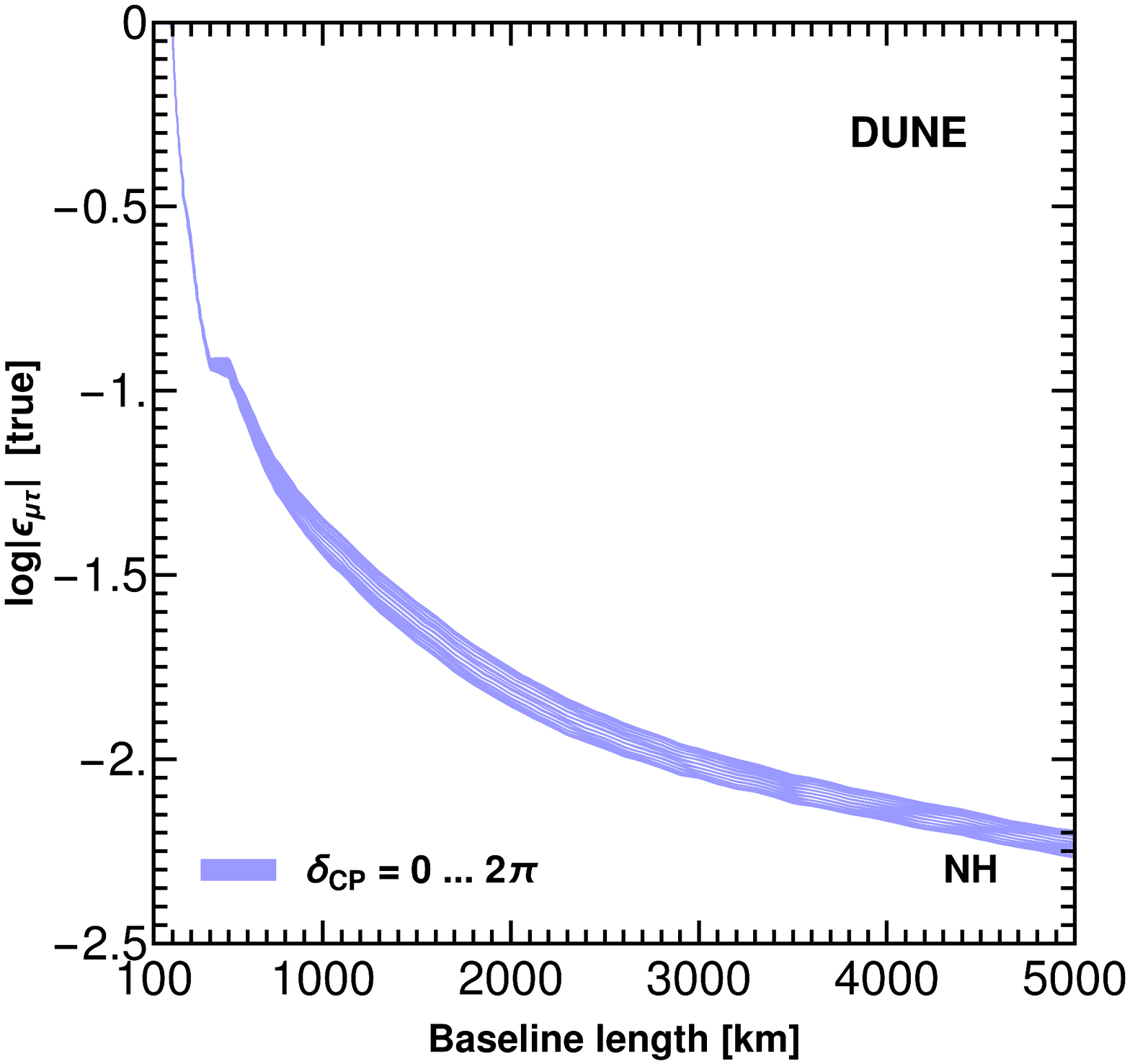}
\end{subfigure}
\begin{subfigure}{0.49\textwidth}
\includegraphics[width=6.0cm]{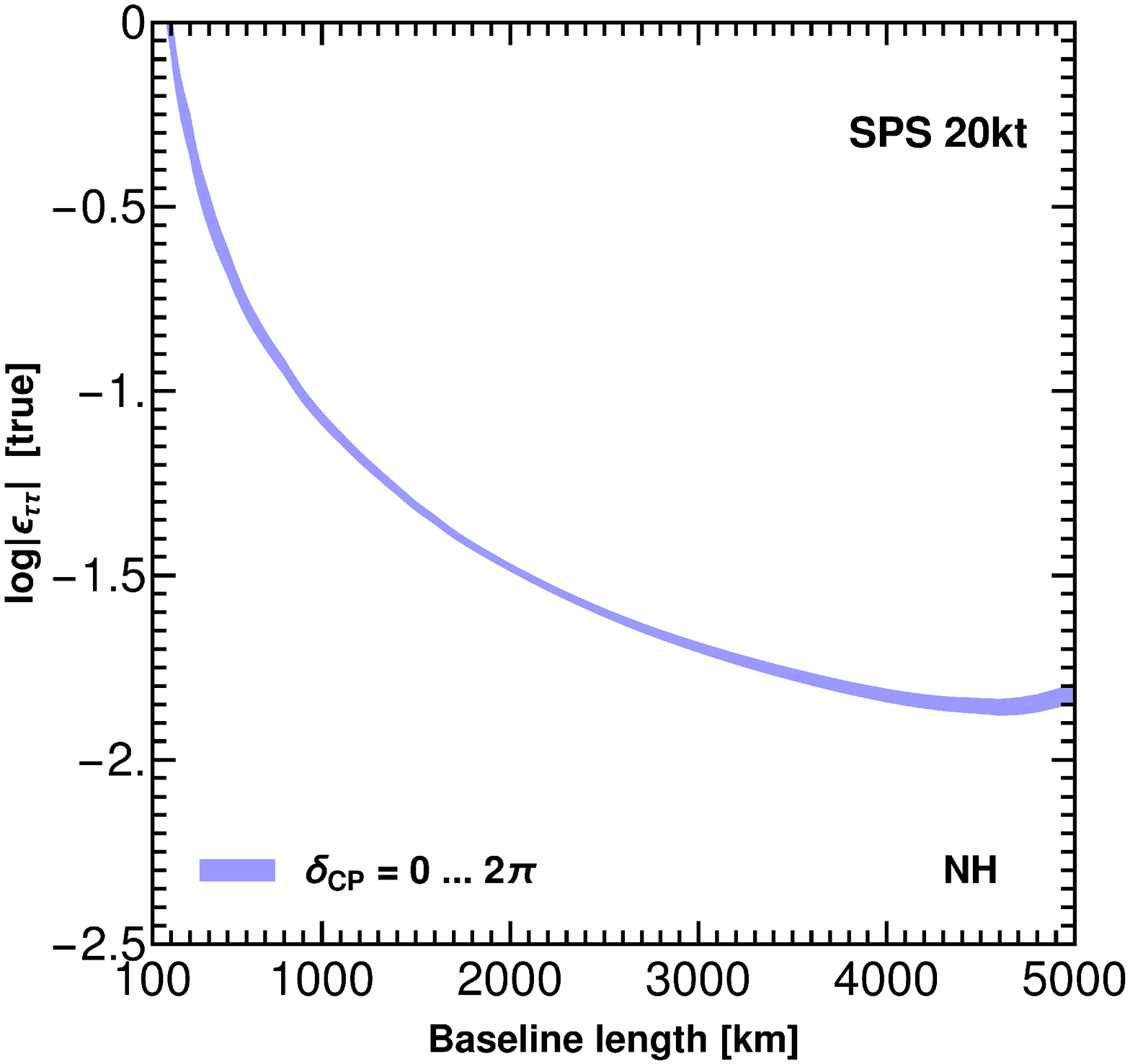}
\end{subfigure}
\begin{subfigure}{0.49\textwidth}
\centering
\includegraphics[width=6.0cm]{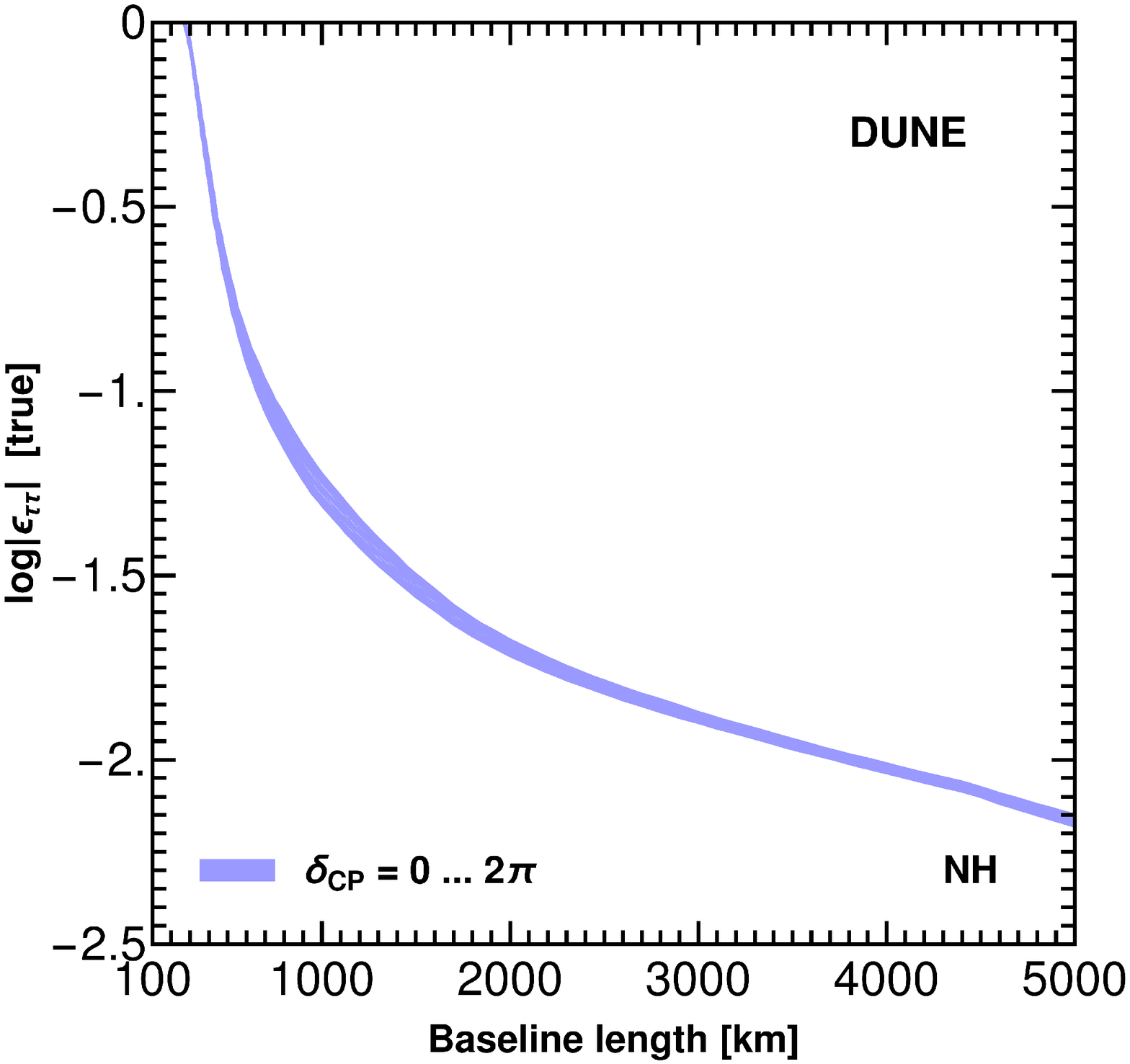}
\end{subfigure}
\caption{\label{fig3}As in Figure~\ref{fig2}, but for $|\varepsilon^m_{\mu\tau}|$ and $|\varepsilon^m_{\tau\tau}|$. For these parameters the correlation effects with $\delta_{CP}$ are tiny, which allows a more exact estimate for the upper bounds of these parameters.}
\end{figure}

We determine the upper bounds for $\varepsilon_{\alpha\beta}^m$ by evaluating the mNSI discovery potential, that is, the sensitivity to rule out SI in favor of mNSI. The non-observation of mNSI then allows
us to set new 90$\%$ confidence limits (CL). The non-observation of NSI then allows to set new  limits for $\varepsilon_{\alpha\beta}^m$. For each $\delta_{CP}$ value, 90 $\%$ CL contour is found and the results merged in a contour band. The bands in $(L,\varepsilon^m_{\alpha\beta})$-plane are plotted in Figures~\ref{fig2} and \ref{fig3}. We find a strong interference from $\delta_{CP}$ to $|\varepsilon^m_{e\mu}|$ and $|\varepsilon^m_{e\tau}|$, and a small interference to all other mNSI parameters.

As expected, the discovery potential is increased when baseline increases, even though the effect shrinks when the baseline is long to begin with. For \textbf{DUNE}, the interference from $\delta_{CP}$ uncertainty is magnified compared to \textbf{SPS}. Changing the hierarchy in all cases and benchmarks produces slightly tighter constraints.

\section*{Acknowledgments}

The author expresses his gratitude to the Magnus Ehrnrooth foundation for financial support and to the organizers of 18th Lomonosov Conference on Elementary Particle Physics for their invitation and hospitality. In addition, the author is grateful to University of Helsinki awarding Chancellor's travel grant for the author's conference participation.

\end{document}